\documentclass[openacc]{rstransa}

\def\be{\begin{equation}}
\def\ee{\end{equation}}
\def\bea{\begin{eqnarray}}
\def\eea{\end{eqnarray}}

\def\gb{g_\text{bar}}
\def\go{g_\text{obs}}
\newcommand{\dd}{\mathrm{d}}

\usepackage{bm}

\titlehead{Review}

\begin{document}

\title{(Exhaustive) Symbolic Regression and model selection by minimum description length}

\author{
Harry Desmond$^{1}$}

\address{$^{1}$Institute of Cosmology \& Gravitation, University of Portsmouth, Dennis Sciama Building,\\ Portsmouth, PO1 3FX, United Kingdom}

\subject{machine learning, astrophysics}


\corres{Harry Desmond\\
\email{harry.desmond@port.ac.uk}}

\begin{abstract}
Symbolic regression is the machine learning method for learning functions from data. After a brief overview of the symbolic regression landscape, I will describe the two main challenges that traditional algorithms face: they have an unknown (and likely significant) probability of failing to find any given good function, and they suffer from ambiguity and poorly-justified assumptions in their function-selection procedure. To address these I propose an exhaustive search and model selection by the minimum description length principle, which allows accuracy and complexity to be directly traded off by measuring each in units of information. I showcase the resulting publicly available Exhaustive Symbolic Regression algorithm on three open problems in astrophysics: the expansion history of the universe, the effective behaviour of gravity in galaxies and the potential of the inflaton field. In each case the algorithm identifies many functions superior to the literature standards. This general purpose methodology should find widespread utility in science and beyond.
\end{abstract}


\begin{fmtext}

\section{Introduction}\label{sec:intro}

A key activity in science is summarising observational data with functional fits. Either one wants a ``fitting function'' with which to propagate some correlation into another part of the analysis, or one wishes to learn the ``law'' that governs the data. Traditionally these two tasks have been tackled in different ways. The creation of fitting functions is normally done ``by eye'': one plots the data and estimates the types of operators and their composition that may give a good fit. This is supplemented with a trial-and-error step: if a given func-
\end{fmtext}
\maketitle
\noindent tional form does not give quite the right asymptotic behaviour, for example, another one is tried and the process iterated until a satisfactory fit is achieved. Such a procedure has on occasion been implemented also for learning ``scientific laws'', the most notable example being Planck's (and his antecedents') discovery of (parts of) the blackbody function. More often however, the discovery of laws is achieved primarily by theory. Partially or completely independently of data, the theorist proposes principles or hypotheses that lead to certain functional relations between observables. These functions are then tested on the data to assess the theory's veracity.

This traditional approach has drawbacks. How can one be sure that the fitting functions one creates are in any sense optimal? One way is for the function to find a theoretical underpinning, as happened with the blackbody formula through quantum mechanics. But purely empirically there is no guarantee that any particular aspect of the function that may be important in applications of it---say its interpolation or extrapolation behaviour---is robust unless one has a quantitative assessment of the function's quality relative to others. Even were such a quality metric available, assessing (all?) other possible functions seems infeasible. The same concerns apply to physical laws extracted empirically: the true features of the law may not be captured by an imperfect function generation procedure. The top-down approach (creating functions in an extra-empirical way before bringing in the data) of course has different concerns: in that case one must assess the reliability of theoretical arguments rather than a regression procedure. Might there be a way to greatly expand the capacity to learn laws directly from data, mitigating or even obviating these concerns?

Enter symbolic regression (SR), the machine learning (ML) framework for extracting functions directly from data. SR aims to automate and perfect the process described above. To explain it, it is useful to begin with the better-known form of regression, which I term \emph{regular} or \emph{numerical}. Here one specifies a priori the functional form that one wants to fit and the regression is only over the numerical values of the function's free parameters. SR generalises this procedure by bringing the functional form itself---the operators and their ordering---into the search space. This substantially increases the difficulty: not only is the search space much enlarged, but the lack of continuity between operators invalidates the use of techniques such as gradient descent that form the staple of parameter optimisation. But there are also clear advantages, most obviously that in regular regression the functional form one imposes is likely to be suboptimal or just plain wrong. SR removes confirmation bias because the user is not required to make that important decision.

While regular regression has a long and venerable history, the advent of ML has introduced some additional ``competitors'' to SR. These are methods like neural networks, Gaussian processes and random forests which can capture the correlations present in a dataset to high fidelity by compounding very many simple functions. These methods excel at ``mindless'' regression and classification tasks: they can produce accurate predictions within the domain of their training set, but very rarely produce scientific insight into a system. This makes them best suited to summarising correlations that one wishes to treat entirely as ``nuisance'', either because the underlying physics is already thoroughly well-known (e.g. the emergent behaviour of ``baryonic physics'' in galaxy evolution), or because it is not believed to be possible to learn basic science from them. SR, on the other hand, excels in cases where one \emph{does} care about the functional form of a relation, perhaps because one believes it to reflect the meaningful physics governing the system. When successful, SR uncovers either the true equation that generated the data---if there is one---or else simply the best possible functional representation. That said, SR can also be valuable for constructing emulators without any demand for, or benefit to be gained from, interpretability: such symbolic emulators are highly portable (e.g. they do not require the neural network emulator's constrained weights and biases), rapid to evaluate and potentially very accurate (e.g.~\cite{syren}).

Exploring the parameter space is quite different when this includes operators. The most common method for generating trial functions is called a genetic algorithm (GA), which works by analogy with natural selection \cite{turing, David, haupt}. This begins by creating a population of functions with typically random examples. One then calculates the ``fitness'' of each one, which is its accuracy on the target dataset, and kills off the functions that fail some fitness cut. Then the next generation of functions is produced through \emph{mutation} and \emph{crossover}. A mutation changes one parts of a function, either a single operator or connected set of operators. A crossover mixes parts of two functions by swapping branches to produce two new functions. This new generation is then assessed against a stopping criterion---for example whether a function of sufficient accuracy has been produced, or a time limit---and if the criterion is not met the process is repeated in the hope of producing ever-better functions. Popular algorithms in this category are Operon \cite{operon}, PySR \cite{cranmer2020discovering} and DataModeler (a proprietary Mathematica add-on; \cite{DM}).

This review focuses on an alternative, non-stochastic approach to SR. A particular emphasis will be the \emph{model selection metric}, i.e. the quantity used to determine how good trial functions are. I will begin by describing the traditional way in which functions are evaluated (Sec.~\ref{sec:assessing}). I will then describe the approach of Exhaustive Symbolic Regression, which overhauls both the function generation and function assessment methodology (Sec.~\ref{sec:esr}). Emphasis here is on the \emph{minimum description length} selection metric (Sec.~\ref{sec:mdl}). I will then describe three astrophysical applications of ESR+MDL, showcasing its power on the expansion rate of the universe (Sec.~5\ref{sec:app1}), galaxy dynamics (Sec.~5\ref{sec:app2}) and cosmic inflation (Sec.~5\ref{sec:app3}). I then describe future developments and conclude. Throughout the article $\log$ has base $e$.

\section{Traditional function assessment}\label{sec:assessing}

Suppose we have a set of trial functions (e.g. from a GA)---how should we score them? Let us consider again the analogy with regular regression. The analogues of candidate functions in this case are points in the space of the pre-defined function's free parameters. In a Bayesian context these are scored by two metrics: the likelihood they give the data, and their prior probability. These multiply by Bayes' theorem, normalised by the evidence, to form the posterior. The best solution is the one that maximises the posterior, with an uncertainty given by a confidence interval of the posterior probability distribution. In a frequentist context one would use just the likelihood.

In SR, focusing on the likelihood or posterior produces an immediate problem. Now that we are allowed to vary operators we can produce arbitrarily complex functions, which typically means that they can be made arbitrarily accurate on the dataset in question. As an example, consider that one can perfectly fit any set of $N$ datapoints with a polynomial of degree $N-1$. But such extremely complex functions that maximise the likelihood are likely severely overfitted and hence extrapolate or generalise very poorly. This means that some measure of \emph{simplicity} needs to be included in the function selection procedure: one wants some optimal trade-off between simplicity and accuracy.

The simplest way of doing this is (unfortunately) the approach traditionally taken in SR; this is to include simplicity as an incommensurate second objective in the regression. This is quantified with a ``complexity'' heuristic, for example the number of nodes in the function's tree representation (i.e. the number of operators, parameters or variables in the function). Functions are then plotted on the 2D plane of accuracy (measured by maximum likelihood or posterior value, or its poor-man's version mean square error) and complexity, producing a \emph{Pareto} (two-objective) optimisation problem. There is a privileged set of functions on the Pareto plane, which are the most accurate for their complexity. These functions---one at each complexity value---form the \emph{Pareto front} and are termed ``Pareto optimal'': any other function has higher inaccuracy and/or complexity and is ``Pareto dominated'' by the optimum functions.

From this perspective all functions along the Pareto front are ``the best'', and they cannot be compared. No metric is provided for how much gain in accuracy is required for an increment of complexity to be warranted. A second heuristic is therefore required for deciding where along the Pareto front to select the best function(s) from. One could eyeball the functions and pick the one deemed to be most attractive, one could decide one wants a function of a certain complexity, one could score the functions on both a training and test dataset and look for the complexity at which the accuracy of the latter starts to become significantly worse than that of the former (indicative of overfitting), or one could stipulate some function of accuracy and complexity to determine the final score. These procedures have no rigorous justification and are up to the user, yet may radically alter the outcome of the regression. We will see in the next section that a superior method exists.

\section{Exhaustive Symbolic Regression}\label{sec:esr}

\subsection{Motivation and Operation}\label{sec:esr_operation}

The traditional approach to SR outlined above faces two major challenges. First, a stochastic search has some completely unknown probability of failing to find any given good function. This may not be a serious issue if that probability is small in practice, but we will see in Sec.~3\ref{sec:benchmarking} that this is not the case.
The failure probability is a function of the algorithm's hyperparameters, but one does not know in advance how to set those to maximise efficacy for the dataset in question. This makes SR results untrustworthy, as there may be any number of better-performing functions than those found. Second, judging equations on their Pareto-optimality depends crucially on the definitions of accuracy and complexity. Changing these will move functions around on the accuracy--complexity plane and change the ones that lie on the Pareto front. Most algorithms adopt mean square error (MSE) as the accuracy measure, but this only accurately describes the data likelihood if the uncertainties on the data points are Gaussian and constant. More importantly the complexity definition is largely arbitrary: some approaches use the number of operators, parameters and variables in the function, some use the depth of the function’s tree representation, while others adopt behavioural rather than structural measures like the degree to which the functions are nonlinear \cite{Order_nonlinearity}. The incommensurability of accuracy and complexity then necessitates another unmotivated heuristic.

\emph{Exhaustive Symbolic Regression} (ESR; \cite{ESR}) is designed to overcome these problems. The first is solved by searching function space exhaustively, guaranteeing discovery of each and every good function, and the second by replacing complexity heuristics with a precise measure of the information content of the function called its description length.

To do an exhaustive search ESR generates every single possible function from some basis set of operators up to a maximum complexity, where complexity is defined as the number of nodes in the function's tree representation.
The operator basis set and maximum complexity are the only things that must be specified by the user, although the maximum complexity is typically set by the computational resources available.
Generating all functions involves generating all possible tree templates, where the nodes are labelled by their arity (number of arguments), and decorating the trees by considering all permutations of the operators in the basis set with the correct arity. We then simplify the functions and remove duplicates using a set of function-comparison rules (\textit{tree reordering, parameter permutations, simplifications, reparametrisation invariance, parameter combinations}). This establishes the \emph{unique} functions, which are all inequivalent to each other and are representatives of sets of behaviourally identical but structurally different functions (e.g. $\theta x$ and $x/\theta$, where $\theta$ is a free parameter). Finally, we find the maximum-likelihood values of the free parameters appearing in the unique functions through nonlinear optimisation, and broadcast these results to all other members of the equivalent sets using the Jacobians of the transformations that relate them. Although the maximum likelihood values of all functions in such a set must be identical, they may possess different description lengths (see below), and hence our search for the lowest description length function ranges over these variants as well as the unique functions. Full details may be found in \cite{ESR}.

This is a computationally expensive procedure due to the huge number of possible functions at higher complexities. One's computational budget therefore limits the complexity one can reach. A typical limit is complexity 10, for which a full ESR run takes $\sim$200 CPU-hours (the scaling with complexity is exponential, so even greatly enhanced computational resources could not extend the maximum complexity by $\gtrsim1$). This depends on the operator basis set (the more operators, the more functions at given complexity and hence the lower the maximum achievable complexity), and is necessarily approximate because the procedure is imperfectly parallelisable. Although the parameter fitting is embarrassingly parallelisable (each function is treated completely separately to all the others), the simplification steps must compare functions and hence cannot be done in isolation. Note that only the parameter fitting is dataset-specific: the function generation and simplification depend only on the operator set, allowing the user to benefit from publicly available pre-computed function sets \cite{esr_zenodo}. This reduces runtime by more than a factor of 2. For reference, straight lines and power-laws have complexity 5, while the Navarro--Frenck--White (NFW; \cite{NFW}) function describing halo density profiles ($\theta_0 (x(x+\theta_1)^{\theta_2})^{-1}$) has complexity 9: any function not much more complex is within scope of ESR.
The full ESR code is publicly available.\footnote{\url{https://github.com/DeaglanBartlett/esr}}

\subsection{Benchmarking}\label{sec:benchmarking}

To illustrate the advantage of a guaranteed search, this section demonstrates the unreliability of stochastic algorithms. We take perhaps the simplest benchmark dataset (\texttt{feynman\_I\_6\_2a}) from the \textit{Penn Machine Learning Benchmarks} dataset, as used in the \textit{SRBench} competition \cite{SR_review}. This comprises $10^5$ datapoints generated from an unknown univariate function without scatter. In addition to ESR we ran five state-of-the-art SR algorithms on the data: PySR, DataModeler \cite{DM}, FFX \cite{FFX}, QLattice \cite{QLattice} and Operon. The test was conducted under exam conditions, with each algorithm given equal opportunity (full details in \cite{ESR}).

\begin{figure*}[h!]
\centering\includegraphics[width=\linewidth]{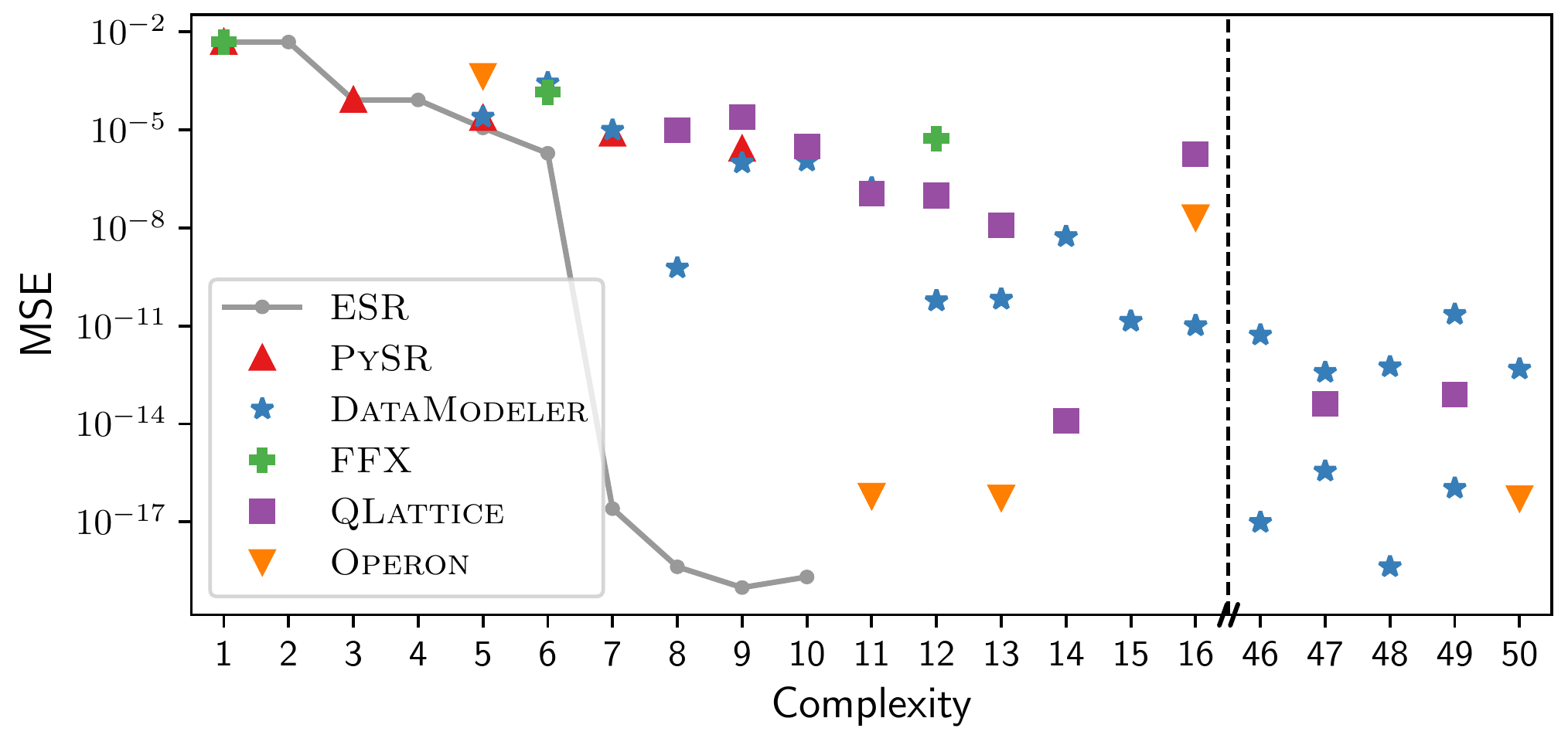}
\caption{Pareto front of mean square error against function complexity produced by six SR algorithms on the \texttt{feynman\_I\_6\_2a} benchmark dataset. Only ESR and Operon identify that the data is drawn from a standard normal, and only ESR finds this function in its simplest form.}
\label{fig:benchmark}
\end{figure*}

Fig.~\ref{fig:benchmark} shows the Pareto front of MSE against complexity returned by each algorithm. The most noticeable feature is the cliff in MSE produced by ESR at complexity 7, which indicates a substantial improvement in the functional fit not seen by any other algorithm. The others can discover the most accurate functions up to complexity 5, but beyond that find only marginally improved solutions. On closer inspection it is seen that not only has ESR produced the best-fitting function, but in fact it has achieved the holy grail of SR which is to find the true function that generated the data. The best complexity-7 function is $y=\theta_1 \theta_0^{x^2}$, where $\theta_0 = 0.6065$ and $\theta_1 = 0.3989$. This is remarkably similar to $\theta_0 = 1/\sqrt{e}$ and $\theta_1 = 1/\sqrt{2\pi}$, suggesting that the data were drawn from a standard normal distribution. Inputting these exact values for the parameters yields an MSE $=3\times10^{-33}$, which is 0 to within machine precision. That ESR did not achieve this directly is due to its numerical tolerance in the parameter optimisation, which also explains why slightly different MSE values are produced for the variants of the standard normal at complexities 8-10. The other SR algorithms gave no indication of this true structure, but simply produced approximate fitting functions. The exception to this is Operon, which does in fact produce a standard normal albeit overparametrised so that it appears at complexity 11.

The conclusion is that most SR algorithms fail even on very simple problems.

\section{The minimum description length principle}\label{sec:mdl}

With ESR we are guaranteed not to miss any function (built from the user-defined basis operator set and up to the maximum complexity), but the problem of ranking them remains. Placing them on the accuracy--complexity plane is sure to construct the true Pareto front (as in Fig.~\ref{fig:benchmark}), but this does not address the arbitrariness of complexity measure or incommensurability of complexity and accuracy. We need a metric that can put accuracy and complexity on the same footing to produce an objective, one-dimensional ranking.

Such an upgrade is analogous to that from maximum-likelihood to Bayesian inference. Consider deciding between two models for some data. Simply considering the maximum-likelihood value, the more complex model is likely to be preferred, a procedure with no safeguard against overfitting. But the model selection metric of Bayesian inference---the \emph{Bayesian evidence}---naturally penalises the more complex model; this model has lower prior values over the region of parameter space that the data prefers, so is less predictive. This is reflected in approximations such as the Bayesian information criterion (BIC), which trade off accuracy (maximum-likelihood value $\hat{\mathcal{L}}$), with the number of free parameters $p$. The addition of another parameter must increase the likelihood by some minimum amount for it to be warranted by the data. A heuristic measure such as the PySR score for combining accuracy and complexity \cite{pysr} is akin to making up a function for trading off $\hat{L}$ and $p$ in ignorance of the evidence.

How can we generalise this argument to include also a penalisation for more complex \emph{structures} of functions, as well as more (and more finely-tuned) parameters as assessed by the evidence? The answer is quite simple (perhaps the reader can already guess), but we will find it instructive to come at it from the new angle of \emph{information theory}.

Why do we want a functional fit in the first place? From an information-theoretic perspective it is to provide a more efficient representation of the data: instead of communicating the full dataset directly, one can hope to convey almost as much by communicating the function instead, which requires many fewer bits. In the limiting case in which the function fits the data perfectly this is completely lossless; otherwise one must either accept a loss of information or supplement the communication with the residuals of the data around the function's expectation, allowing the data to be fully reconstructed. This is formalised in the \emph{minimum description length} (MDL) principle, which states that the goodness of each functional fit is quantified by the number of nats (log-$e$ bits) of information needed to convey the data with the help of the function, which is called the description length \cite{rissanen, MDL_review_1}. The best function minimises this.

This requires communicating the function, or hypothesis $H$, and the residuals $D|H$ where $D$ denotes the data, and may be written $L(D) = L(H) + L(D|H)$,
where $L$ is the description length operator. ($L(D|H)$ may alternatively be considered the information loss, which acts as a penalisation term.) How many nats are needed for each piece? Under an optimal coding scheme called Shannon--Fano \cite{cover_thomas}, the residual term is simply the minimum negative log-likelihood, describing the function’s accuracy on the data: $L(D|H) = -\log(\hat{\mathcal{L}})$. $L(H)$ comprises the information content of the operators in the function and the values of any free parameters it contains. If the function's tree contains $k$ nodes (what we are calling the function's complexity) populated by $n$ distinct operators, there are $n^k$ possible permutations corresponding to $\log(n^k)=k\log(n)$ nats of information.
Note that the value of this depends on the operator basis set: for example it is smaller for tan(x) (if tan is included in the operator set) than sin(x)/cos(x) (if only sin, cos and $\div$ are included). Any natural numbers $c_j$ arising as part of the function simplification process are encoded in $\sum_j \log(c_j)$ further nats.

Finally we must encode the values of the maximum-likelihood parameters, $\hat{\theta}_i$. This is done by choosing a precision $\Delta_i$ with which to record the $i^\text{th}$ parameter, thus representing it as $|\theta_i|/\Delta_i$ increments encoded in $\log(|\theta_i|/\Delta_i)$ nats.
The larger $\Delta_i$, the fewer nats are required to convey $|\theta_i|/\Delta_i$ but the worse the hit to the likelihood when $\theta_i$ is rounded to the nearest $\Delta_i$. We choose the $\Delta_i$ that optimises this trade-off to produce the lowest total description length. This is found to be at $\Delta_i = (12/I_{ii})^{1/2}$, where $I$ is the observed Fisher information matrix at the maximum-likelihood point,
$I_{ij} = -\left. \frac{\partial^2 \log\mathcal{L}}{\partial\theta_i \partial\theta_j} \right|_{\hat{\theta}}$.
The sign of the parameter must be communicated with a further $\log(2)$ nats. Putting this all together produces our final formula for the description length \cite{ESR}:
\be\label{eq:DL}
L(D) = -\log(\mathcal{L}(\hat{\mathbf{\theta}})) + k\log(n) - \frac{p}{2} \log(3) + \sum_j \log(c_j) + \sum_i^p \left(\frac{1}{2}\log(\mathsf{I}_{ii}) + \log(|\hat{\theta}_i|)\right),
\ee
This is the quantity with which we score functions, lower being better.

We seem to have gone a long way from the Bayesian evidence discussed at the start of the section. In fact we have unknowingly come full circle. Consider the generalisation of the evidence $\mathcal{Z} \left( D | f_i \right)$ for the probability of function $f_i$ given the data, including explicitly a \emph{prior $P(f_i)$ on the function itself}:
\begin{equation}\label{eq:evidence}
    P \left(f_i | D \right)
    = \frac{1}{P(D)} \int
    P \left( D | f_i, \bm{\theta}_i \right)
    P \left( \bm{\theta}_i | f_i \right)
    P \left( f_i \right)
    \dd \bm{\theta}_i \\
    \equiv \frac{ P \left( f_i \right)}{P(D)} \mathcal{Z} \left( D | f_i \right).
\end{equation}
Up to the overall normalisation term $P(D)$ (the probability of getting the data from any function), this simply multiplies the standard evidence by the function prior. Generalising the implementation of Occam's razor with regard to parametric complexity, it seems clear that in the Bayesian context the penalisation of \emph{structural complexity} must derive from this functional prior term. Just as introducing additional parameters is apt to lower the prior over the high-likelihood region of numerical parameter space, introducing additional operators should lower the prior over the high-likelihood region of functional parameter space. The more operators one has, the less likely any given combination. This is the ``quite simple'' solution alluded to above.

How does this relate to $L(D)$? To see the connection, calculate $\mathcal{Z}$ under the \emph{Laplace approximation}, which treats the posterior as a multidimensional Gaussian around the maximum-posterior point:
\begin{equation}\label{eq:laplace}
    - \log P \left(f_i | D \right) \simeq - \log P \left( f_i \right) - \log\mathcal{Z} \left( D | f_i \right)
    \simeq - \log P \left( f_i \right) - \log \mathcal{\hat{H}} - \frac{p}{2} \log 2\pi + \frac{1}{2} \log \det \hat{\bm{I}}^{\mathcal{H}},
\end{equation}
where the first relation neglects the overall normalising constant $P(D)$ and the second implements the Laplace approximation. $\mathcal{H}$ denotes the posterior and $\hat{\bm{I}}^{\mathcal{H}}$
is the observed Fisher information matrix for the posterior at the maximum posterior point.
Eqs.~\ref{eq:laplace} and~\ref{eq:DL} now look suspiciously similar: both have terms depending on the maximum-likelihood value, the number of free parameters and the observed Fisher information.
It is simple to show \cite{priors} that the two equations are effectively equivalent provided
\begin{equation}\label{eq:MDL_prior}
    - \log P \left(f_i \right) =  k \log n  + \sum_\alpha \log c_\alpha.
\end{equation}
In other words, the description length is nothing more than (the negative log-likelihood of) the Bayesian probability for the model under a specific choice of prior that penalises more structurally complex functions.

This parallel also reveals a key freedom we have in calculating $L(D)$. In Bayesian statistics priors are subjective, representing our degree of belief before measurements are made. We may therefore consider replacing the functional prior $k\log(n) + \sum_j \log(c_j)$ by something else. Why would we want to do that? One reason could be that that prior is
completely ignorant of the structure of successful functions in the domain in question, for example those that describe ``more physical'' solutions. As an example, suppose one is considering an oscillatory system and wishes to assess whether some response goes as $\sin(x_1) + \sin(x_2)$ or $\sin(\sin(x_1+x_2))$. These possess exactly the same operators, and hence have identical values of $k\log(n)$. However, our background physics knowledge tells us that oscillatory systems are frequently described by sums of sines but very rarely nested sines, so it is intuitively reasonable that our prior should favour the former function.

This may be formalised by adapting an algorithm from language processing called the \emph{Katz back-off model} \cite{Katz_1987}. This takes in a training set of equations describing successful functions within some pertinent domain, from which it learns probabilities of operator combinations. A hyperparameter controls the length of these combinations, and another the minimum number of occurrences of the combination in the training set for its probability to be directly calculated; for combinations that do not appear that many times, the algorithm ``backs off'' to instead consider a combination of one shorter length. This method therefore models prior probabilities for SR-generated functions according to their similarity to those in the training set, the hope being 1) that that set successfully describes something about the problem in question, and 2) that that success is reflected in the prevalence of various operator combinations and hence may be learnt by the model. We have constructed an implementation of the Katz model which may readily be used as a drop-in replacement for the $k\log(n) + \sum_j \log(c_j)$ prior \cite{priors},\footnote{\url{https://github.com/DeaglanBartlett/katz}} where it is also shown to perform well on a variety of benchmark tests.

\section{Astrophysical applications}\label{sec:apps}

We want SR whenever we want a functional form for some relation. Sometimes, theory (established or speculative) predicts a function and we want to know how good it is. At other times, there is no prediction but we want either empirical inspiration for the governing law or a fitting formula to summarise the phenomenology. This section describes three cases in the former category, where concordance cosmology, gravity theory or inflationary model building makes a prediction in astrophysics. We wish to see how well it performs relative to all other possible predictions, and whether we could have recovered the theory directly from the empirical data.

\subsection{The cosmic expansion rate}\label{sec:app1}

Cosmology studies the overall structure and evolution of the universe. The standard cosmological model is called $\Lambda$CDM because it posits that the present-day universe consists primarily of dark energy ($\Lambda$) and cold dark matter (CDM). $\Lambda$CDM makes predictions called the \emph{Friedmann equations}
for the expansion rate $H$ of the universe as a function of time.
They are written not in terms of time directly, which is unobservable in cosmology, but rather in terms of redshift $z$, which describes the amount by which photons' wavelengths have been stretched due to expansion of the universe en route to us. This acts as a proxy for time, with higher redshift meaning further into the past.

The simplest Friedmann equation, assuming matter is pressureless, is
\begin{equation}\label{eq:f1}
H^2(z) = H_0^2 (\Omega_\Lambda + \Omega_m (1+z)^3),
\end{equation}
where $H_0\equiv H(0)$ is the current expansion rate and $\Omega_\Lambda$ and $\Omega_m$ are dimensionless quantities describing the relative prevalence of dark energy and dark matter respectively.
This equation can be fitted to $H(z)$ data and its parameters constrained. We could even calculate the goodness-of-fit relative to alternative functions, for example with different equations of state for dark energy or dark matter, through the Bayesian evidence.
But this procedure necessarily specifies the models under consideration a priori and can only rank them against each other; how can we see how good the model is \emph{relative to all possible others}?

SR offers the promise of determining the optimal form of $H(z)$ directly from the data, without requiring theorists to first come up with trial functions. In particular, \emph{E}SR produces the complete list of all possible (simple) $H(z)$ functions and scores them on the data, producing an objective ranking from which one can see how any given model fares relative to all others. This is the closest one can come to absolute goodness-of-fit testing in a Bayesian context.

We use two datasets to do this \cite{ESR}. The first, simpler, datasets involved \emph{cosmic chronometers} (CCs), stellar populations in galaxies that evolve passively such that their relative ages can be determined from spectroscopy. Combined with the difference in their measured redshifts, this affords a determination of $\Delta z/\Delta t$, a discrete approximation to $-(1+z) H(z)$. This dataset contains 32 points and derives from \cite{Moresco_2022}.
The second utilises \emph{Type Ia Supernovae} (SNe), which are explosions of massive stars and the end of their lives. SNe are ``standardisable candles'', meaning that, after some corrections relating to properties of their lightcurves, their luminosity is universal. This enables calculation of their luminosity distances, from which may be calculated $H(z)$. Here we adopt 1590 SNe from the Pantheon+ sample, including the covariance matrix between SNe which modulates the likelihood
\cite{Scolnic_2021}. We simply plug these datasets separately into ESR (using the $k\log(n)$ function prior) and turn the crank to produce all possible $H(z)$ functions and their DLs. We use the operator basis set $\{x\equiv1+z, \theta, \text{inv}, +, -, \times, \div, \text{pow}\}$.

\begin{figure*}[h!]
\centering\includegraphics[width=0.48\linewidth]{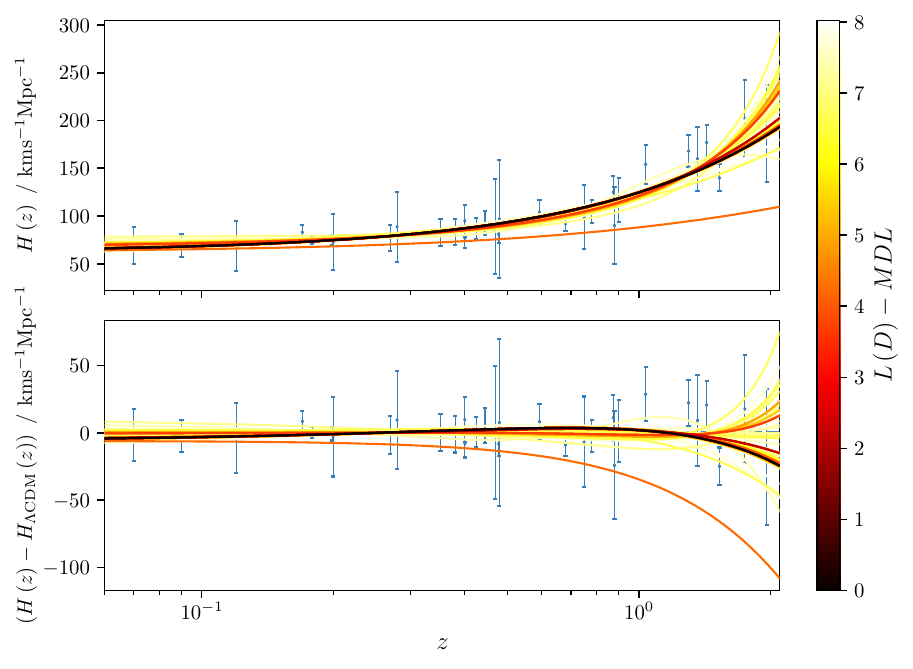}
\centering\includegraphics[width=0.51\linewidth]{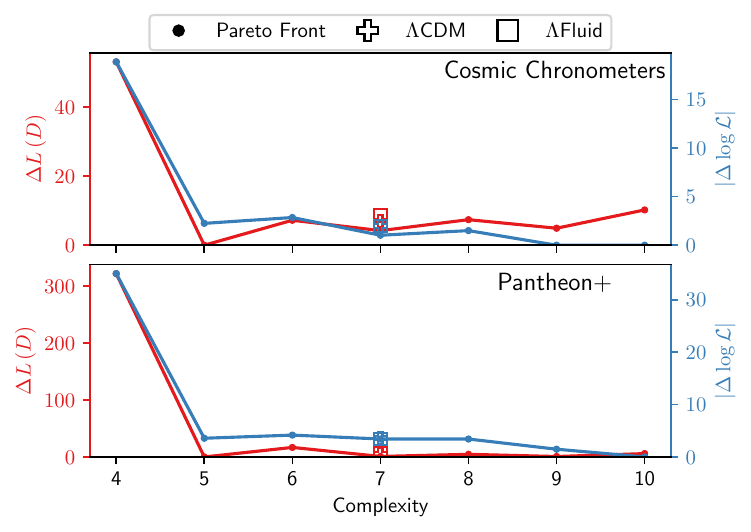}
\caption{\emph{Left:} Top 150 ESR $H(z)$ functions overplotted on the CC data (upper panel), and the residuals from Eq.~\ref{eq:f1} (lower panel). \emph{Right:} Pareto fronts for both datasets with the literature standards (Friedmann equations) shown as separate symbols. Reproduced from \cite{ESR}.}
\label{fig:hubble}
\end{figure*}

The results are shown in Fig.~\ref{fig:hubble}: the left panel shows the 150 best functions found by ESR colour-coded by their description length relative to the best function, compared to the CC data, along with their residuals relative to Eq.~\ref{eq:f1} in the lower panel. The right panel shows the best DL and $\log\mathcal{L}$ discovered at each complexity in both datasets compared to the Friedmann equations. ``$\Lambda$fluid'' is a generalisation of Eq.~\ref{eq:f1} that allows matter to have pressure. For both datasets we find that the MDL function has complexity 5, lower than the Friedmann equations at complexity 7. These MDL functions are
\begin{equation}\label{eq:Hz_best}
H^2(z) = \theta_0 (1+z)^2 \:\:\:\:\: \text{(CCs)}\:, \:\:\:\:\:\:\: H^2(z) = \theta_0 (1+z)^{1+z} \:\:\:\:\: \text{(SNe)},
\end{equation}
and are preferred over the simple Friedmann equation (Eq.~\ref{eq:f1}) by 7.12 and 4.91 nats for the CC and SN data respectively, which corresponds to probability ratios of 1240 and 136. We find 38 functions better than Eq.~\ref{eq:f1} for CCs and 36 for SNe. Note that while the Pareto front as traditionally defined in terms of likelihood continues falling to high complexity, $L(D)$ has a well-defined minimum beyond which the small gain in accuracy is swamped by the loss in simplicity. This informs the user whether sufficiently complex functions have been considered: if not, the minimum in $L(D)$ will not yet have been seen.

What are we to make of the sub-optimality of the Friedmann equations? It is premature to conclude that they are ruled out as the equations governing the data. It will be noticed that both functions in Eq.~\ref{eq:Hz_best} produce the same Taylor expansion for $H(z)$ as the Friedmann equation up to second order in $z$, making their behaviour (hence likelihood) very similar over the limited redshift range of the data. They are however simpler functions, with smaller complexity terms in their description lengths. It is indeed a feature of the MDL procedure that with limited quality or quantity of data, the likelihood term is less important than the complexity terms and hence simpler functions will be preferred. More data is therefore needed to settle definitively on the best function(s), and the MDL formalism can tell us precisely what are the properties of the data required \cite{ESR}.

\subsection{The radial acceleration relation}\label{sec:app2}

From the late-time expansion rate of the universe we turn to the internal dynamics of galaxies \cite{sr-rar}. A puzzle here has been the enduring success of an alternative to the dark matter hypothesis on which $\Lambda$CDM is based. This theory---Modified Newtonian Dynamics (MOND) \cite{Milgrom_1,Milgrom_2,Milgrom_3}---posits that the ``missing gravity problem'' in galaxies (e.g. that rotation curves become flat at large galactocentric radius rather than declining in a Keplerian fashion) is due not to missing mass, but rather to an alteration in either objects' gravity or inertia at exceedingly low accelerations below a new universal parameter $a_0 \approx 1.2\times10^{-10}$ m/s$^2$. This takes the form
\be\label{eq:mond}
\vec{g}_\text{obs} = \nu(g_\text{bar}/a_0) \: \vec{g}_\text{bar},
\ee
Where $\gb$ is the acceleration sourced by the visible stars and gas in galaxies (``baryons'') and $\go$ is the dynamical acceleration that drives objects' motions.
$\nu(x)$---the ``interpolating function'' (IF)---can take any form provided it satisfies the limits $\nu(x)\rightarrow1$ for $x\gg1$ and $\nu(x)\rightarrow x^{-1/2}$ for $x\ll1$. These are called the ``Newtonian'' and ``deep-MOND'' limits; the former recovers Newtonian dynamics at $a\gg a_0$ (e.g. in the Solar System), while the latter is the modification required to get flat rotation curves. Classic IF choices are the ``Simple'' ($\nu(x) = 1/2 + (1/4 + 1/x)^{1/2}$)
and ``RAR'' ($\nu(x) = 1/(1-\exp(-\sqrt{x}))$) functions.

This bizarre theory has astonishing successes and atrocious failures (see \cite{moriond} for a recent review). Its principal success is in predicting, and explaining the properties of, the ``radial acceleration relation'' (RAR), which directly correlates $\gb$ and $\go$ through measurements of late-type galaxies' HI and optical photometry and rotation curves through spectroscopy \cite{RAR}.
This relation has been found to possess a minute scatter and no correlated deviations from MOND predictions \cite{uRAR}, to satisfy four stringent criteria for ``fundamentality'' in galaxies' radial dynamics \cite{fRAR}, and to evince a continuation to much lower accelerations through weak lensing \cite{mistele}.
We wish to use SR to determine 1) whether the classic MOND IFs are optimal descriptions of the RAR data, and 2) if not, whether the optimal functions possess the correct limits to be considered new IFs. This will help determine the extent to which the RAR supports MOND.

\begin{figure*}[h!]
\centering\includegraphics[width=0.535\linewidth]{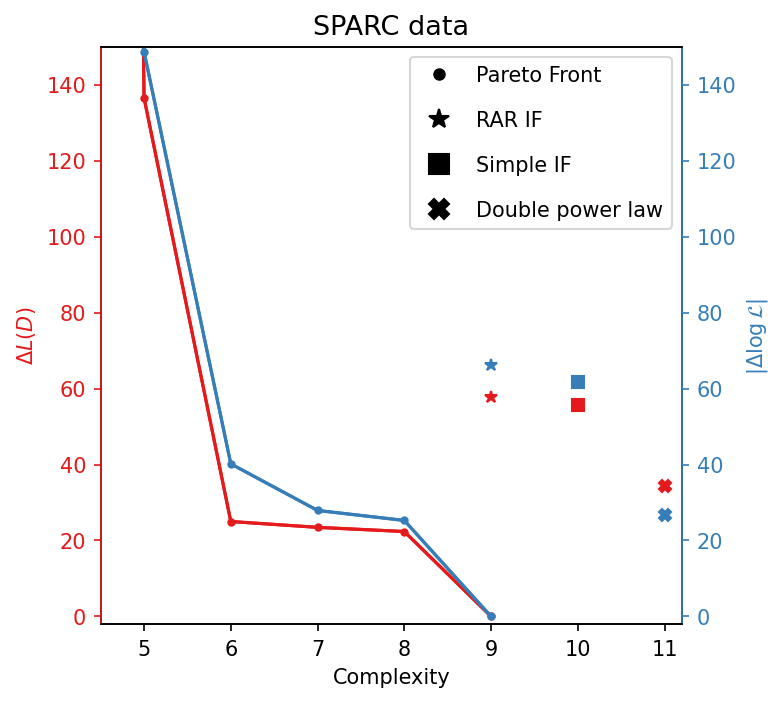}
\centering\includegraphics[width=0.45\linewidth]{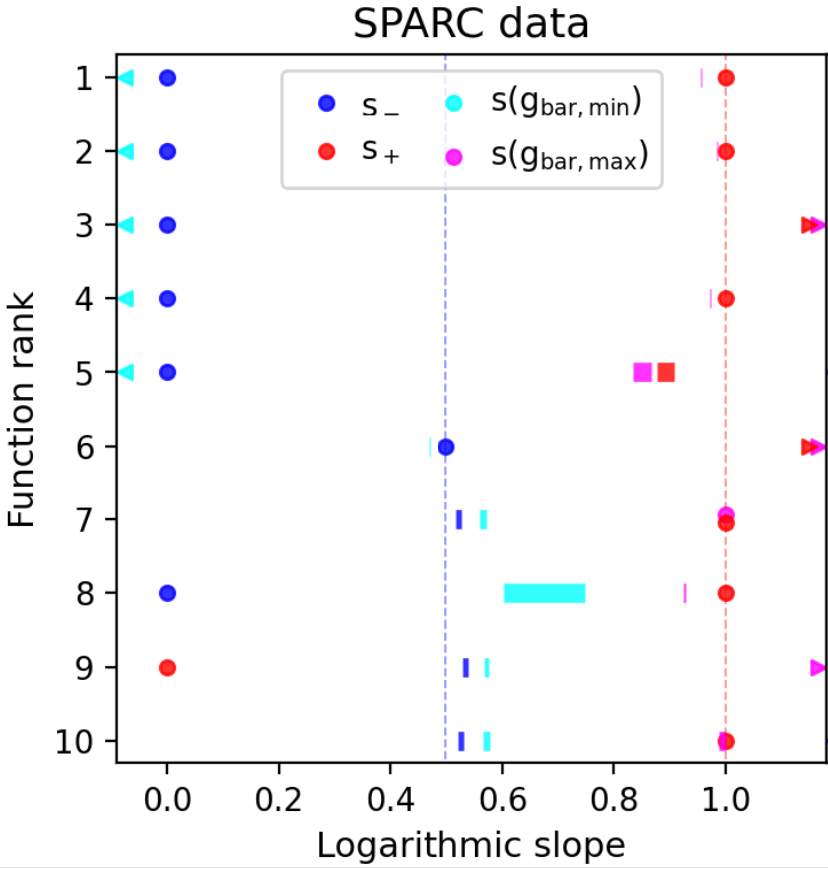}
\caption{\emph{Left:} Pareto front identified by ESR for the RAR compared to the three most common literature fits.  \emph{Right:}  Logarithmic slopes of the 10 best ESR functions in the limit $\gb\rightarrow0$ (blue) and $\gb\rightarrow\inf$ (red), and at the lower (cyan) and upper (magenta) edges of the data. Where these depend on free parameters of the function they are shown as bands indicating the 95\% C.L. The MOND lower and upper slopes are shown by vertical dashed lines; especially the lower slope is rarely satisfied by the best functions. Reproduced from \cite{sr-rar}.}
\label{fig:rar}
\end{figure*}

We adopt the operator basis set \{$\gb, \theta, \exp, \text{sqrt}, \text{square}, \text{inv}, +, -, \times, \div, \text{pow}$\}, and once again simply plug the data into ESR using the $k\log(n)$ function prior to create and score the full set of simple $\go(\gb)$ functions. The results are shown in Fig.~\ref{fig:rar}.  We find many functions better than the MOND IFs. While most of these possess a Newtonian limit, very few possess a deep-MOND limit: more commonly $\go$ tends to a constant at low $\gb$. Indeed, the fact that the probability of a function is $\propto \exp{-L(D)}$ (Sec.~\ref{sec:mdl}) lets us compute a probability \emph{over all functions} for certain limiting behaviour, which in this case is $\sim$1 for $\go\rightarrow$ const as $\gb\rightarrow0$. However, this is found also to be true on mock data generated using the RAR IF which explicitly possesses such a limit \cite{sr-rar}. Thus, while the best functions are not MONDian, one would not expect them to be even if MOND did in fact generate the data, given the data's limited dynamic range and sizeable uncertainties. This is similar to the situation in Sec.~5\ref{sec:app1}, where the data could not have been expected to return unambiguously the Friedmann equation even if it did in fact generate the data. Better measurements are therefore required to know whether or not the RAR significantly evidences MOND in its functional form.

\subsection{Inflation}\label{sec:app3}

For our final application we move to the very early universe \cite{inflation}. An largely-accepted add-on to $\Lambda$CDM postulates that within $\sim10^{-32}$s of the Big Bang the universe underwent a period of accelerated expansion known as \emph{inflation}. This caused the universe to grow by a factor $\sim e^{60}$, smoothing it out and producing the seeds of structure growth through quantum fluctuations. Inflation is driven by one or more scalar fields, called \emph{inflatons}. The simplest scenario is single-field, slow-roll inflation in which a single inflaton $\phi$ rolls down a shallow potential gradient, generating a de Sitter-like epoch manifesting exponential expansion. This has a simple Lagrangian and equation of motion
\be
\mathcal{L} = -\frac{1}{2} g^{\mu\nu} \partial_\mu \phi \partial_\nu \phi - V(\phi) \: \: \Rightarrow \: \: \ddot{\phi} + 3 H \dot{\phi} + V'(\phi) = 0.
\ee
The unknown in these equations is the potential $V(\phi)$: the task for SR is to find the best such functions.

Part of the problem with inflation---and the reason it is so amenable to SR---is that it is highly underdetermined: there is a dazzling variety of theories of inflation \cite{encyc}, but effectively only three numbers with which to test them. These are the amplitude and tilt of the primordial power spectrum and upper limit on the tensor-to-scalar ratio, all measured from the cosmic microwave background by the \textit{Planck} satellite \cite{planck1,planck2}:
\be
A_s = (0.027 \pm 0.0027) M_\text{pl}, \:\:\: n_s = 0.9649 \pm 0.0042, \:\:\: r < 0.028 \: (95\% \: \text{C.L.}).
\ee
This is the data with which we infer $V(\phi)$ using ESR. To explore the effect of operator basis set we consider two possibilities: Set A: $\{\phi, \theta, \text{inv}, \exp, \log, \sin, \text{sqrt}, \text{square}, \text{cube}, +, -, \times, \div\}$ and Set B: $\{\phi, \theta, \text{inv}, \exp, \log, +, -, \times, \div, \text{pow}\}$. Set A favours powers that are simply composed of sqrt, square and cube, while set B increases flexibility with the general $\text{pow}$ operator.

\begin{figure*}[h!]
\centering\includegraphics[width=0.565\linewidth]{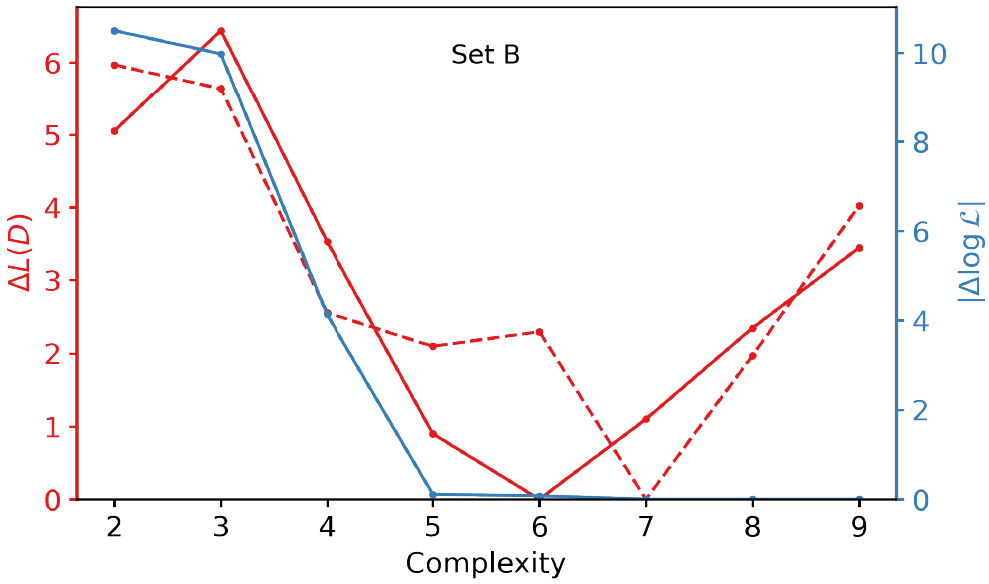}
\centering\includegraphics[width=0.425\linewidth]{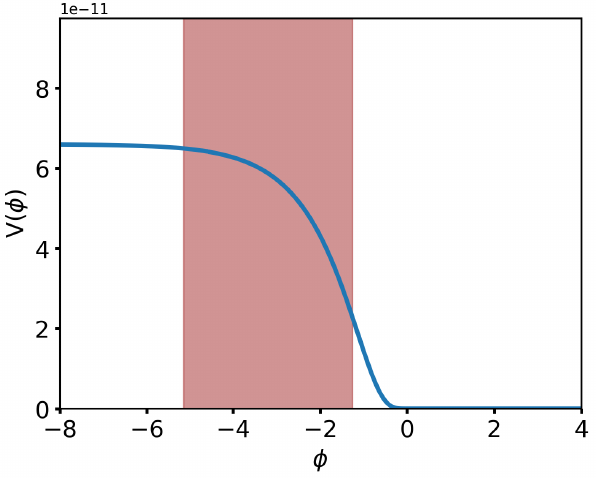}
\caption{\emph{Left:} Pareto front for inflaton potentials using operator basis set B, using fiducial MDL (solid) or Katz back-off model (dashed) function priors. \emph{Right:} The potential favoured by the $k\log(n)$ prior, $\exp(-\exp(\exp(\exp(\phi))))$, with the slow-roll region over which inflation occurs shown by the red shaded band.}
\label{fig:inflation}
\end{figure*}

The Pareto front for basis set B is shown in Fig.~\ref{fig:inflation} (left), using both the $k\log(n)$ (solid) and Katz (dashed) function priors. For both basis sets the MDL function, at complexity 6, is $\exp(-\exp(\exp(\exp(\phi))))$, which, despite its byzantine structure, produces a very well-behaved slow-roll region (Fig.~\ref{fig:inflation}, right). It is however unlikely to derive from a sensible underlying particle physics theory of inflation. We therefore consider replacing the $k\log(n)$ prior, favouring structurally simpler functions, with the Katz prior presented in Sec.~\ref{sec:mdl}, which favours functions more similar to those in a training set. We choose as training set the \textit{Encyclopaedia Inflationaris}, containing almost all known literature theories of inflation and their corresponding inflaton potentials \cite{encyc}. The hope is that the ``physicality'' of these potentials (that is, their derivability from a well-behaved particle physics theories) is reflected in the structure of their operator combinations and may therefore be picked up by the Katz model. With this prior, the MDL functions are $\theta_0 (\theta_1 + \log(\phi)^2)$ for operator set A, and $\theta_0 \phi^{\theta_1/\phi}$ for set B. These are more reasonable functions that may have a physical underpinning.

We can also use the full ranking afforded by ESR to compare the best functions to literature standards. As an example, for operator set B with the $k\log(n)$ prior, the Starobinsky, quadratic and quartic models place at ranks 1272, 8697 and 10839 respectively. Clearly, as in the previous applications, SR is required to find the best functions.

\section{Upcoming developments and future work}\label{sec:future}

SR is an exciting emergent field with much potential for the future. On a $\sim$10-year timescale it is set to become a key method of the physicist's (and scientist's) toolbox, not only for finding fitting functions in a principled and automated way but also for learning physical laws directly from data. The symbolic side of ML will begin to see the successes and widespread application that the ``numerical'' side already has.

We are planning a range of upgrades and further applications of ESR to realise this potential. The most significant is a comprehensive set of upgrades to the ESR algorithm itself. This involves a rewrite in Julia for automatic differentiation, expediting the parameter optimisation and description length calculation, and uses a faster and more efficient scheme for simplifying functions and removing duplicates. This will raise the complexity cap from 10 to $\sim$13. ESR 2.0 will also enable the modelling of multiple independent variables, and of snapping parameters to integers, rational numbers and fundamental constants (e.g. $c$, $e$, $\pi$) where that would reduce the description length. We are also planning ways to combine the virtues of ESR at low complexity and GAs at higher complexity. This could include using the complete characterisation of function space afforded by ESR to initialise GAs with the best-performing low-complexity functions, or to optimise their hyperparameters through knowledge of the relations between such functions. The MDL principle is independent of ESR and already finding use in GAs, where it has been shown to improve performance \cite{Bogdan_MDL}.

On the practical side we have only scratched the surface of applications in astrophysics, let alone science more generally. Our next priority is to use ESR to learn the optimal functional forms of halo density profiles, both from $N$-body simulations and observations of galaxy kinematics. This is a case where the standard, NFW profile was identified using the ``by-eye'' method described in Sec.~\ref{sec:intro}, and hence has scant justification despite being a crucial component of many astrophysical and cosmological analyses. Similarly the so-called halo mass function and galaxy mass and luminosity functions, describing the abundance of galaxies and dark matter halos as a function of their luminosities and masses. The standard here is the Schechter function \cite{Schechter}, which is also highly likely to be improvable by a rigorous SR procedure. Indeed, any of the multifold fitting functions prevalent in astrophysics may be upgraded by applying SR, and ESR provides a simple and effective method for doing so.

\section{Summary \& Conclusion}\label{sec:conc}

Symbolic regression is the machine learning method that codifies and automates the practice of empirical science, namely the creation of functions and equations describing data. While this is traditionally done ``by eye'', SR is a task for computers---affording an enormous increase in processing power---if effective methods can be constructed.

SR is traditionally achieved by stochastic algorithms (e.g. genetic programming), with candidate functions assessed by locating the Pareto front in accuracy and complexity and then applying an additional ad hoc rule for selecting a single ``best'' function. I have argued that such methods face two serious challenges: 1) they have a significant probability of failing to find any given good function, and 2) the model selection procedure is unfounded in its arbitrary definition of complexity and unjustified heuristic for breaking the Pareto front degeneracy. To overcome the first I propose \emph{Exhaustive Symbolic Regression} (ESR), and to overcome the second I propose the \emph{Minimum Description Length} (MDL) principle. ESR implements an efficient algorithm for generating and optimising the parameters of \emph{all} functions composed of a user-defined basis set of operators up to a maximum complexity, guaranteeing discovery of all good solutions. MDL measures functions' quality with an information-theory-motivated metric that makes accuracy and simplicity commensurable, affording a principled one-dimensional ranking. It is essentially the Bayesian evidence plus a prior on functions that penalises those containing more, and more varied, operators. An alternative prior, based on a \emph{Katz back-off} language model, instead penalises functions with combinations of operators that are rarer in a training set of equations.

I showcase ESR+MDL on three hot topics in astrophysics: the late-time expansion rate of the universe, the effective behaviour of gravity in galaxies and the potential of the field driving inflation. In each case, ESR discovers functions considerably superior to literature standards (the Friedmann equation, MOND and Starobinsky, quadratic and quartic inflation respectively), illustrating its ability to uncover effective symbolic representations of data purely empirically. This bodes well for future discovery not only of optimal fitting functions, but, more ambitiously, of physical laws directly from data.

SR is only just starting to take off. The near future will see an extensive overhaul of the ESR algorithm, greatly improving its efficiency and capabilities. Synergy with genetic algorithms is also promising, combining their advantages in different regimes of complexity. At the same time there is a range of astrophysical (and other) fitting functions ripe for improvement.
I suspect the golden age of symbolic machine learning not to be far away.

\vskip6pt

\ack{I thank Deaglan Bartlett, Pedro Ferreira, Gabriel Kronberger, Lukas Kammerer and Tomas Sousa for the collaborations on which this work is based. I am supported by a Royal Society University Research Fellowship (grant no. 211046).}

\bibliographystyle{RS}
\bibliography{desmond}

\end{document}